# Giant spin oscillations in an ultracold Fermi sea

J. S. Krauser[1,†], U. Ebling[2,†], N. Fläschner[1], J. Heinze[1], K. Sengstock[1,3,*], M. Lewenstein[2,4], A. Eckardt[5], C. Becker[1,3]

[1]ILP - Institut für Laserphysik, Universität Hamburg, Luruper Chaussee 149, 22761 Hamburg, Germany

[2]ICFO - Institut de Ciències Fotòniques, Av. Carl Friedrich Gauss 3, 08860 Castelldefels, Spain

[3]ZOQ - Zentrum für Optische Quantentechnologien, Universität Hamburg, Luruper Chaussee 149, 22761 Hamburg, Germany

[4]ICREA - Institució Catalana de Recerca i Estudis Avançats, Lluís Companys 23, 08010 Barcelona, Spain

[5]MPI-PKS - Max-Planck-Institut für Physik komplexer Systeme, Nöthnitzer Str. 38, 01187 Dresden, Germany

[*]To whom correspondence should be addressed. E-mail: klaus.sengstock@physnet.uni-hamburg.de

[†]These authors contributed equally to this work.

**Collective behavior in many-body systems is the origin of many fascinating phenomena in nature ranging from swarms of birds and modeling of human behavior to fundamental magnetic properties of solids. We report on the first observation of collective spin dynamics in an ultracold Fermi sea with large spin: We observe long-lived and large-amplitude coherent spin oscillations, driven by local spin interactions. At ultralow temperatures, Pauli blocking stabilizes the collective behavior and the Fermi sea behaves as a single entity in spin space. With increasing temperature, we observe a stronger damping associated with particle-hole excitations. As a striking feature, we find a high-density regime where excited spin configurations are collisionally stabilized.**

A fundamental question in many-body physics is how the collective behavior of many microscopically-interacting particles gives rise to macroscopic properties such as superconductivity, giant magnetoresistance or superfluidity despite the suppression of Bose-Einstein condensation in liquid Helium. Ultracold quantum gases have proven to be excellent candidates to address questions like these owing to their unrivaled control over all crucial experimental parameters like the interaction strength and the dimensionality of the system. Collective behavior of ultracold fermions is of particularly high interest due to its analogy to other systems in nature like electrons in solids, neutron stars or baryonic matter. Recent experiments studied spin ½ fermions e.g. the BEC-BCS crossover (*1,2*), Mott physics (*3,4*), thermodynamic and transport properties (*5-7*), collective excitations (*8,9*) and magnetic ordering (*10,11*). Beyond this, ultracold atoms offer the striking possibility to realize fermionic high-spin systems with more than two spin components, which constitute a completely new class of many-body systems. Here, collective effects involving the spin degree of freedom may lead to fascinating interaction-induced phenomena (*12-19*). For bosonic atoms with a total spin of $f=1,2,3$ spin-dependent phenomena have been intensively studied (*20,21* and references therein). In these experiments, the macroscopic occupation of a common single-particle wave function, which is inherent to Bose-Einstein condensation, strongly favors collective behavior. For fermions, in contrast, every single-particle state is occupied with only one particle and hence macroscopically many of these states are occupied in a Fermi sea. This naturally raises the following questions: Can the whole Fermi sea exhibit collective and coherent dynamics of its internal spin degree of freedom (Fig. 1A)? Does this dynamics preserve the spatial structure or do emerging spin components populate new spatial modes? Recently, first experiments with high-spin fermions allowed to study ground-state and dynamical properties in the regime of deep optical lattices (*22,23*), few-body physics (*24*) and also the emergence of spatial spin waves (*25*), leaving, however, the questions raised above largely unanswered. Here, we report on the first observation of coherent spin dynamics in an ultracold Fermi sea. Our detailed studies reveal three fundamental findings: (i) Giant collective spin oscillations lasting for seconds at ultralow temperatures where spatial degrees of freedom are effectively frozen out. (ii) A continuous suppression of the collective behavior for increasing temperatures. (iii) A high-density regime where excited spin configurations surprisingly remain stable for seconds which we attribute to a collision-induced stabilization mechanism.

We have observed these effects in a quantum degenerate gas of $^{40}$K in an optical dipole trap (*26*), well isolated from its environment. The hyperfine ground state of $^{40}$K has a total spin $f=9/2$, which gives rise to ten spin states with magnetic quantum numbers $m=-9/2,\ldots,9/2$. To record the population of these different spin components, we release the atoms from the trap and subsequently measure the spin occupations after expansion in a Stern-Gerlach field (*26*). To understand collective spin-changing dynamics, we first consider the underlying microscopic collisions, well described by s-wave scattering: Two fermionic atoms collide and change their spin configuration ($m_1, m_2 \to m_3, m_4$), conserving the total magnetization ($m_1+m_2=m_3+m_4$) and obeying the Pauli exclusion principle ($m_1 \neq m_2$ and $m_3 \neq m_4$). The interplay between different quadratic Zeeman energies and differential spin-dependent interaction energies for the different spin configurations determines whether spin oscillations can occur or not (20). At large magnetic fields, spin-changing collisions are suppressed energetically and the Fermi sea remains in its initial spin configuration. For low magnetic fields, the two energy scales become comparable and resonant spin oscillations are induced. As a key result, Fig. 1B demonstrates this kind of coherent spin oscillations for timescales as long as $2\,\mathrm{s}$ for an ultracold atomic Fermi ensemble. In the following we show that this peculiar behavior can be theoretically described via a mean-field approach (25). Furthermore, all key observations are explained surprisingly very well within a single-spatial-mode approximation, i.e. the whole Fermi sea evolves in time as a single spin entity.

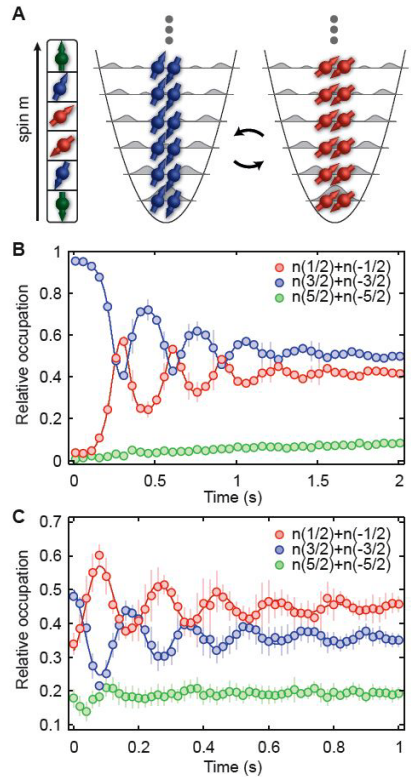

**Fig. 1. Collective and coherent spin-changing oscillations in a Fermi sea.**

**A** Sketch of fermionic many-body spin-changing dynamics. **B** Collective spin-changing dynamics with small seed. Plotted are the relative populations $n(m)$ of different spin states as a function of time. Solid lines are guides-to-the-eye. The magnetic field, peak density and temperature are $B = 0.1\,\mathrm{G}$, $n_\mathrm{p} = 5.9 \times 10^{12}\,\mathrm{cm}^{-3}$ and $T = 0.14\,\mathrm{T_f}$. **C** Collective spin-changing oscillations of a Fermi sea prepared with coherences (26). Plotted is $n(m)$ as a function of time. Solid lines are fits to the data, from which we extract the oscillation amplitude and frequency (26). The experimental parameters are $B = 0.12\,\mathrm{G}$, $n_\mathrm{p} = 5.9 \times 10^{12}\,\mathrm{cm}^{-3}$ and $T = 0.13\,\mathrm{T_f}$.

The high-spin Fermi sea is described by the single-particle density matrix in Wigner representation $w_{ij}(\mathbf{x},\mathbf{p})$, being a matrix in spin space depending on position and momentum. Its time evolution obeys the Boltzmann equation (26)

$$\partial_t w_{ij}(\mathbf{x},\mathbf{p}) = \partial_0 w_{ij}(\mathbf{x},\mathbf{p}) + \left[V(\mathbf{x}) + qS_z^2, w(\mathbf{x},\mathbf{p})\right]_{ij} + I_{ij}(\mathbf{x},\mathbf{p}). \tag{1}$$

Here, $\partial_0 w_{ij}(\mathbf{x},\mathbf{p})$ accounts for the spin-independent center-of-mass motion of the atoms in the trap and $I_{ij}(\mathbf{x},\mathbf{p})$ is the collisional integral accounting for particle-hole excitations. Spin-changing dynamics is driven by the commutator including the mean-field potential $V_{ij}(\mathbf{x}) = \int d\mathbf{p} \sum_{kl}(U_{klji} - U_{kijl}) \cdot w_{kl}(\mathbf{x},\mathbf{p})$ with the interaction tensor $U_{klji}$ and the quadratic Zeeman shift $qS_z^2$. The commutator drives the spin-changing dynamics in the presence of coherences given by finite off-diagonal matrix elements of $w_{ij}$ (26). In Fig. 1B, the system is prepared with very small initial coherences sufficient to initialize a spin instability followed by collective spin oscillations, a behavior also observed for Bose-Einstein condensates (27).

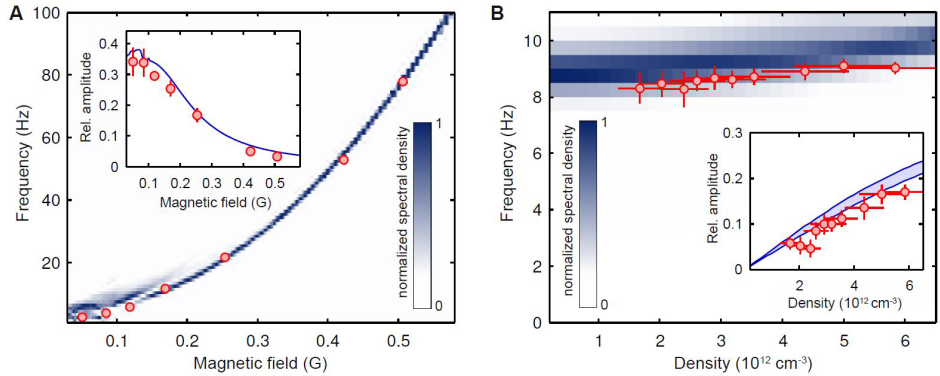

**Fig. 2. Collective behavior at ultralow temperatures.**
Frequency (main graphs) and amplitude (insets) of collective spin oscillations versus magnetic field (**A**) and density (**B**), both for an initial state as in Fig. 1C. Experimental parameters are $n_p = 1.0 \times 10^{13}\,\text{cm}^{-3}$ and $T = 0.22\,T_f$ for **A**, and $B = 0.17\,\text{G}$ and $T = 0.18\,T_f$ for **B**. Data (red points) are deduced from fits (26). Solid lines in insets correspond to the amplitude and the shading in the main graph to a Fourier spectrum, both deduced from calculations within a single-mode approximation without free parameters (26).

The interplay between the kinetic and interaction energy in equation (1) can lead to a fascinating effect: If the spatial dynamics of the trapped atoms is reasonably faster than the spin-changing dynamics, then the spatially-dependent interactions are averaged out leading to an effective long-range interaction potential (*28*). This induces collective spin-changing behavior of the whole Fermi ensemble despite the underlying complex spatial structure (*26*), justifying the above-mentioned single-mode approximation. Fig. 1C shows coherent spin oscillations at ultralow temperatures for an ensemble initially prepared in a particular coherent spin superposition (*26*). In contrast to the instability-induced oscillations shown in Fig. 1B, this state immediately exhibits spin oscillations and can thus be systematically compared to theoretical calculations. At high magnetic fields, we find low-amplitude and high-frequency oscillations in contrast to slow and giant spin oscillations at low magnetic fields as depicted in Fig. 2A. By varying the density and thereby the interaction energy within the Fermi sea (Fig. 2B), we find an increase in amplitude and a rather constant frequency for increasing interaction. The system precisely follows the mean-field description within the single-mode approximation without free parameters. As one key result of this work, we therefore conclude that spin-changing collisions can induce collective behavior in a Fermi sea of trapped atoms preserving the spatial structure for long timescales, even for macroscopic systems with several $10^5$ particles and a spatial extension larger than $100\mu m$ .

A fundamental question is how finite temperature affects the observed collective behavior. Fig. 3A shows the strong impact of increasing temperature on the spin-changing dynamics in our experiment, revealing long-lived oscillations at low temperatures and a nearly full suppression of coherent evolution at higher temperatures. Again we find good agreement for the observed frequencies and amplitudes with the single-mode approximation (see Fig. 3B). However, for higher temperatures we observe substantially stronger damping of the coherent oscillations, as depicted in Fig. 3C. We attribute this to an enhanced rate of particle-hole excitations, which are described by the collisional integral $I_{ij}(\boldsymbol{x},\boldsymbol{p})$ in equation (1). These incoherent collisions are characterized by a finite momentum exchange of the colliding particles. In contrast, the collisions inducing collective spin oscillations conserve the individual momenta and are captured by the mean-field potential. At ultralow temperatures, the Fermi statistics does not provide any free momentum states due to the Pauli exclusion principle, hence collisions involving momentum exchange are strongly suppressed (*29*). For increasing temperatures, however, free momentum states become available in the Fermi sea and allow for increased particle-hole excitations. In a relaxation approximation ansatz (30), we calculate the corresponding damping rate without free parameters and find good agreement with the observed damping for temperatures up to $T = 0.4\,T_f$ (see Fig. 3C). For even higher temperatures the coherent oscillations are fully suppressed, which underlines that ultralow temperatures are the key ingredient to enable collective spin-changing dynamics in a Fermi sea.

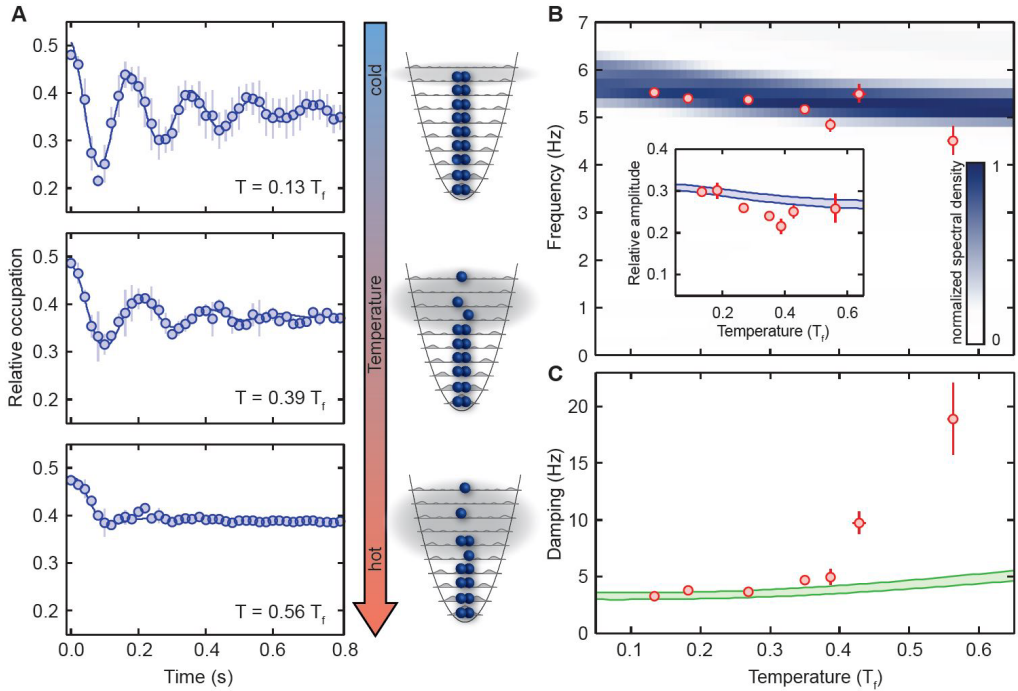

**Fig. 3. Temperature influence on collective spin dynamics.**

**A** Time evolution of the occupations $n(\pm 3/2)$ for three different temperatures (experimental data and fitted curves). **B** Frequency (main graph) and amplitude (inset) for different temperatures. **C** Damping rate versus temperature. The initial state was prepared as for the measurements in Fig. 1C and the experimental parameters are $B = 0.12\,\text{G}$ and $n_p = 5.2 \times 10^{12}\,\text{cm}^{-3}$. Data (red points) are deduced from fits. Amplitude and frequency calculations in **B** are performed within a single-mode approximation, damping calculations in **C** within a relaxation approximation, both without free parameters (26).

Finally, we investigate the occurrence of spin instabilities like the one presented in Fig. 1B as a function of the density and the magnetic field. As a particularly surprising feature we find a regime at high densities where magnetically excited initial states remain stable for seconds. We study this effect by investigating the emergence of spin instabilities, which is depicted in the stability diagram in Fig. 4A. Here, we initially prepare the system in the magnetically excited spin states $m = \pm 3/2$ and map out the $m = \pm 1/2$ occupation after a time evolution of $t = 2\,\mathrm{s}$, where collective spin oscillations have been damped out already and the system is in a quasi-equilibrium configuration. We identify three different regions: ① The Zeeman-protected regime, where spin-changing collisions are off-resonant and the spin configuration remains constant. ② The spin-oscillation regime, where spin-changing collisions are resonant and spin instabilities occur. ③ A stable high-density regime, where the spin configuration does not change, despite the fact that coherent spin-changing oscillations would be resonant, thus possible. To illustrate the behavior in regime ③ more clearly, we show time evolutions at different magnetic fields for two slightly different densities (Figs. 4B,C), revealing an abrupt spin stabilization at small magnetic fields for the larger density. We attribute this stabilization to a self-induced spin-Zeno-effect: At large density, increased incoherent collisions "project" the spins back onto the initial state and thus suppress the coherent evolution observed at low densities. Note that this intriguing spin stabilization is a pure collisional effect, which is highly sensitive to slight changes of the experimental parameters and leads to a completely different macroscopic behavior of the Fermi sea.

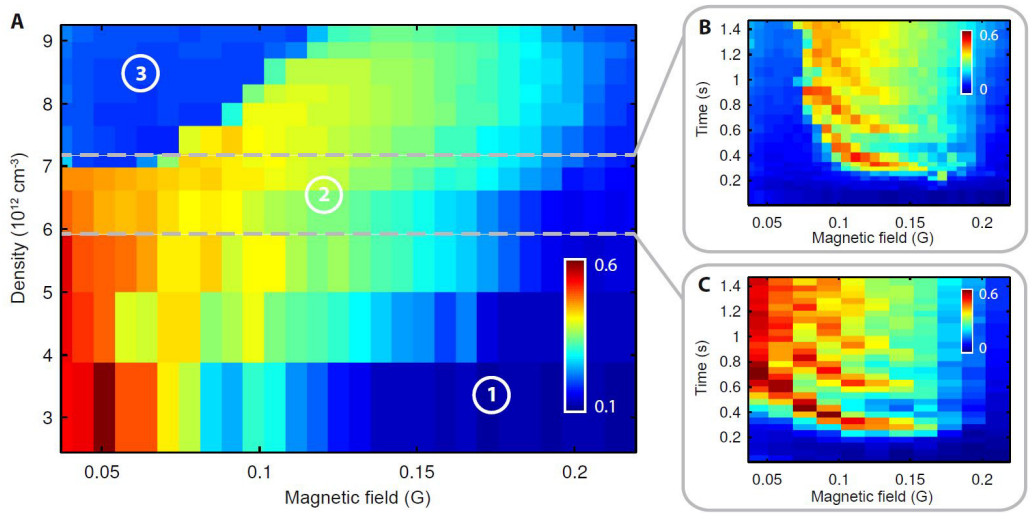

**Fig. 4. Stability diagram of a magnetically-excited Fermi sea.**

**A** Occupation of $n(\pm 1/2)$ after a time evolution of $t = 2\,\mathrm{s}$, dependent on the magnetic field and the density for an initial state as in Fig. 1B. Three regions are identified: a Zeeman-protected regime ①, a spin-oscillation regime ② and a high-density regime ③. **B** and **C** depict the time evolution for different magnetic fields at a density of $n_p = 5.9 \times 10^{12}\,\mathrm{cm}^{-3}$ and $n_p = 7.2 \times 10^{12}\,\mathrm{cm}^{-3}$.

In conclusion, we have studied collective spin-changing dynamics in a degenerate trapped Fermi gas. In particular, we investigated the intriguing interplay between several fundamental processes which either stimulate or suppress the collective behavior and address fundamental open many-body questions. The high experimental control as well as the obtained understanding of our model system open the path to study yet inaccessible exotic phenomena like e.g. the creation of topological structures and textures in superfluid Fermi gases with high spin.

## Acknowledgements

We acknowledge stimulating discussions with N. Cooper, W. Hofstetter, L. Mathey and L. Santos. This work has been funded by DFG grant FOR 801, ERC AdG QUAGATUA, Spanish MICINN (Grant TOQATA (FIS2008-00784), FPI-fellowship) and EU IP SIQS.

# Supplemental Material

## S1 Experimental details

### S1.1 Preparation of an interacting fermionic quantum gas

We sympathetically cool spin-polarized atoms of $^{40}$K in the state $|F=9/2, m=9/2\rangle$ ($F$ denotes the total spin and $m$ the magnetic quantum number) with bosonic $^{87}$Rb to quantum degeneracy, typically realizing samples of $N = 2\times 10^6$ particles at temperatures of $0.1\,\mathrm{T_f}$. Subsequently, we transfer the atoms into a crossed circular-elliptical optical dipole trap operated at a wavelength of $\lambda = 812\,\mathrm{nm}$. Using radio-frequency pulses and sweeps, we prepare an equal mixture of the two hyperfine states $m_1$ and $m_2$ and evaporate this binary mixture by lowering the power of the dipole trap exponentially within a time of $2\,\mathrm{s}$. In the optical dipole trap, we typically realize samples with temperatures of $0.1-0.2\,\mathrm{T_f}$ and particle numbers of $N = 4\times 10^5$. For our experiments, it is necessary to prepare different spin mixtures at very low magnetic fields. Due to the existence of a large number of loss-inducing Feshbach resonances in $^{40}$K, which we measured independently, the realization of many binary mixtures is experimentally challenging. For the experiments presented in Figs. 1C, 2 and 3, we generate an ultracold spin mixture of $m_1, m_2 = \pm 1/2$, which is the magnetic ground state of the system and can be evaporated at $B = 50\,\mathrm{G}$. For the experiments presented in Figs. 1B and 4, we generate an ultracold spin mixture of $m_1, m_2 = \pm 3/2$, which is magnetically excited and cannot be directly evaporated. At a magnetic field of $B = 50\,\mathrm{G}$, we therefore first evaporate a different spin mixture ($m_1 = -1/2$ and $m_2 = 3/2$), which is stable against spin-changing losses due to Pauli blocking. Subsequently, we perform a short radio-frequency sweep of typically $2\,\mathrm{ms}$, transferring atoms from spin state $m_1 = -1/2$ into $-3/2$. After the evaporation, we typically ramp up the dipole trap again for the spin dynamics experiments. The final trapping frequencies in the dipole trap are $\omega_{x,y,z} = 2\pi \times (133, 42, 32)\,\mathrm{Hz}$ for the experiments depicted in Fig. 2A, $\omega_{x,y,z} = 2\pi \times (109, 28, 29)\,\mathrm{Hz}$ for the experiments in Fig. 2B and $\omega_{x,y,z} = 2\pi \times (137, 32, 33)\,\mathrm{Hz}$ for the experiments depicted in Fig. 3 and Fig. 4. We estimate the uncertainty of the trapping frequencies to be $10\%$ and of the magnetic field to be less than $3\,\mathrm{mG}$.

**S1.2 Initialization of collective spin dynamics in a Fermi sea**

A major challenge for observing collective spin dynamics in a fermionic many-body system is to satisfy two preconditions as worked out in section (S2): finite interactions to allow for spin-changing collisions in the sample and finite single-particle coherences. We employ two different approaches to fulfill these conditions and hence to initialize the collective spin dynamics: For the experiments depicted in Figs. 1B and 4, a very small seed in the initial state is sufficient to initialize self-induced collective dynamics. We checked that this instability-induced collective dynamics sets in even if the amount of seed atoms is below the detection limit. For the measurements depicted in Figs. 1C, 2 and 3, in contrast, we explicitly prepare initial coherences as described in the following. Consider the single-particle density matrix, which describes the Fermi sea in its ground-state hyperfine manifold with a total spin of $F = 9/2$ and ten spin states $m = -9/2, ..., 9/2$. The diagonal elements represent the spin occupations, whereas the non-diagonal elements represent the corresponding coherences. In the experiment, we evaporate a binary spin mixture of $m_1, m_2 = \pm 1/2$. The resulting single-particle density matrix $\rho_0$ hence consists of the corresponding two diagonal entries whereas all non-diagonal elements are zero as depicted in Fig. S1A.

Subsequently, we apply a short radio-frequency pulse of about $10\,\mu s$ at a low magnetic field, which creates non-diagonal elements and initializes spin-changing dynamics. At low magnetic field strengths, the radio-frequency coupling is dominated by the linear Zeeman effect, which results in a coupling of all ten spin components.

The resulting density matrix is given by

$$\rho(\vartheta) = \exp(-iS_{x,y} \cdot \vartheta) \cdot \rho_0 \cdot \exp(iS_{x,y} \cdot \vartheta). \tag{S1}$$

Here, $S_{x,y}$ are the corresponding generators of rotation and $\vartheta$ corresponds to the rotation angle in spin space. Fig. S1A shows the impact of the radio-frequency manipulation on the density matrix of the high-spin Fermi sea, where one clearly observes occupations in new spin states as well as emerging coherences. Fig. S1B shows the experimental result of the radio-frequency pulse for different rotation angles corresponding to different radio-frequency pulse powers. We compare the observed spin occupations depending on the rotation angle $\vartheta$ to (S1) and find very good agreement with the Rabi frequency as the only free parameter.

For the experiments shown in Figs. 1C, 2 and 3 we apply a radio-frequency pulse corresponding to a rotation angle in spin space of $\vartheta = 0.44$. This state is depicted in Fig. S1A in (III) and exhibits a nearly equal occupation of the $\pm 1/2$ and $\pm 3/2$ spin components resulting in large-amplitude oscillations which are well suited for our investigations.

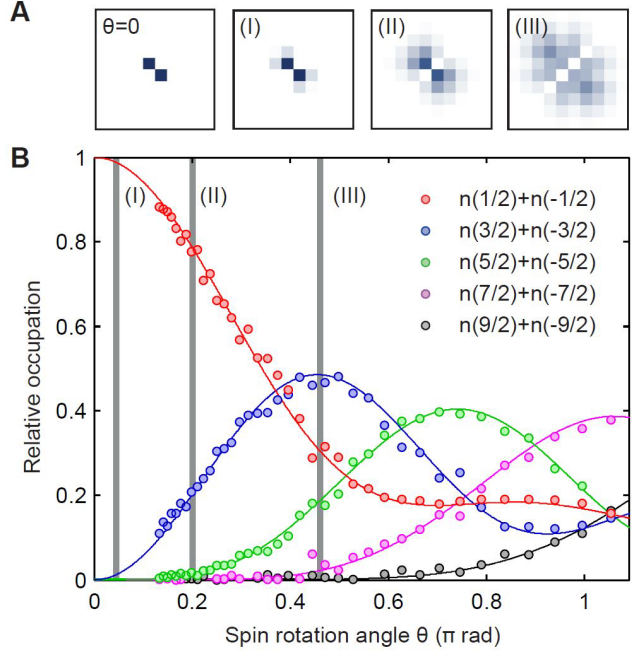

**Fig. S1. Preparation of coherences in a Fermi sea.**

**A** Calculated density matrices for different radio-frequency pulse powers. Diagonal elements represent the spin occupations and non-diagonal elements the corresponding coherences. The elements of the density matrix are ordered from $m = -9/2,...,9/2$. The initial state (left) is the evaporated mixed state with spin states $m_1, m_2 = \pm 1/2$, (I)-(III) correspond to rotation angles $\vartheta$ as indicated in B. The state (III) with $\vartheta = 0.44$ is typically used in the experiments. **B** Measured spin occupations for different rotation angles $\vartheta$. Solid lines are fits resulting from a calculation with equation (S1) with the Rabi frequency as the only free fit parameter. At a magnetic field of $B = 0.5\,\mathrm{G}$, a $100\,\mu\mathrm{s}$ radio-frequency pulse was applied at a frequency of $159\,\mathrm{kHz}$ and for $100\,\mu\mathrm{s}$, applied with different powers onto the atoms in spin states $m_1, m_2 = \pm 1/2$.

### S1.3 Experimental detection analysis

Here, we describe the experimental determination of the spin populations, the total particle number and the temperature of our samples.

In our experiments, we employ time-of-flight imaging. After a variable evolution time, all optical potentials are suddenly switched off and an applied Stern-Gerlach field leads to the separation of the spin states within an expansion time of typically $18.5\,\mathrm{ms}$. We detect the atoms with resonant absorption imaging which allows us to measure the total particle number $N$ as well as all relative spin populations simultaneously.

To determine the temperature of our samples, we release the two-component Fermi gas from the trap and employ a fugacity fit, which is independent of the particular trapping frequencies. We apply this fit to the sample after a time-of-flight of typically $21\,\text{ms}$ without a Stern-Gerlach field and without spin-changing dynamics. The temperature determination is performed separately to avoid deformations of the cloud due to the Stern-Gerlach field. Note that the observed temperatures reflect the initial temperatures of our samples. We checked that applying a short radio-frequency pulse as applied in Figs. 1C, 2 and 3 does not increase the temperature. A slight temperature increase is observed during the experiments which we mainly attribute to single-photon scattering events.

With the measured experimental parameters (the trap frequencies, particle numbers and temperatures), we determine the density and also perform all numerical calculations. In the whole paper, density always refers to the total peak density $n_\text{p}$ of the system. We estimate the errors for the investigated parameters in the following way: Errors in the observed relative spin populations shown in Figs. 1B and C correspond to the standard deviation. The errors of the magnetic field depicted in Fig. 2A reflect the systematic uncertainty of the magnetic field and always lie within the data points. The error in the density shown in Fig. 2B mainly results from uncertainties of the trapping frequencies. Note that in the rather shallow trap configuration employed in this experiment, the influence of the Gaussian shape of the dipole trap is enhanced and might lead to a systematic overestimation of the determined density. Errors in temperature in Fig. 3 are taken from the fugacity fit, which is independent of the particular trapping frequencies, and mostly lie within the data points.

### S1.4 Fitting procedure of the spin oscillations

In this section, we describe how we experimentally determine the frequency, amplitude and damping rate from the observed spin-changing dynamics evaluated in Figs. 2 and 3.

For each experiment, we measure the time-dependent occupation $n^{(m)}(t)$ of the different spin states $m$. Due to our particular preparation, collective spin-changing behavior between the spin configurations $\pm 1/2$ and $\pm 3/2$ is induced. Spin oscillations between other spin configurations are small and do not significantly contribute to the time evolution. We therefore employ the following fitting function:

$$n^{(m)}(t) = O^{(m)} + A \cdot \exp(-\Gamma t) \cdot \cos\left(\omega t + \varphi^{(|m|)}\right).  \tag{S2}$$

Here, $O^{(m)}$ is a general offset for each spin component, accounting for the average occupation. The spin oscillations are described by the second term: $A$ is the amplitude, $\omega$ the frequency, and $\varphi$ the initial phase of the oscillation. We assume a damping rate of the oscillations described by an exponential function with a time constant $\Gamma$. Note that the values for $A$, $\omega$ and $\Gamma$ are extracted in conformity with all participating single-particle spin states, whereas $O^{(m)}$ is fitted independently and $\varphi^{(|m|)}$ is shifted by $90°$ between $\pm 3/2$ and $\pm 1/2$. We also observe a very slow overall increase or decrease of the population of the individual spin states mediated by incoherent spin-changing collisions. This contribution is reproduced by our numerical calculations and very small for the timescales and densities in our experiments. We therefore neglect this feature in the fitting procedure. All given errors of the frequency, the amplitude and the damping rate depicted in Figs. 2 and 3 correspond to fit errors from equation (S2), representing one standard deviation.

## S2 Derivation of the collisionless Boltzmann equation

In this section, we derive the collisionless Boltzmann equation, which we employ in equation (1) in the main text for the description of collective dynamics of a spinor Fermi gas in the mean-field approximation.

In our system, the single-particle state of an atom of mass $m$ is described by a spinor field operator $\hat{\psi}_i(\boldsymbol{x})$, where $i$ indicates the magnetic quantum number. Due to the fermionic statistics, these spinors follow anticommutation rules $\{\hat{\psi}_i^\dagger(\boldsymbol{x}), \hat{\psi}_j(\boldsymbol{y})\} = \delta_{ij}\delta(\boldsymbol{x}-\boldsymbol{y})$. Thus, in second quantization the Hamiltonian is given as

$$\hat{H} = \int d\boldsymbol{x} \sum_{ij} \hat{\psi}_i^\dagger(\boldsymbol{x}) \left[ -\frac{\hbar^2}{2m}\nabla^2 \delta_{ij} + V^{\text{trap}}(\boldsymbol{x}) + q(S_z)_{ij}^2 \right] \hat{\psi}_j(\boldsymbol{x}) + \frac{1}{2}\sum_{ijkl} \int d\boldsymbol{x}\, U_{ijkl} \hat{\psi}_i^\dagger(\boldsymbol{x})\hat{\psi}_k^\dagger(\boldsymbol{x})\hat{\psi}_l(\boldsymbol{x})\hat{\psi}_j(\boldsymbol{x}),$$

consisting of a single-particle and an interaction part. Here and in the following, sums over spin indices include all internal states. The first two terms of the single-particle part describe the motion of atoms in a trapping potential $V^{\text{trap}}(\boldsymbol{x}) = \frac{m}{2}(\omega_x^2 x^2 + \omega_y^2 y^2 + \omega_z^2 z^2)$, whereas the third part accounts for the presence of an external homogeneous magnetic field along the spin quantization axis, generating a non-linear Zeeman splitting. Note that in our system the quadratic Zeeman energy is dominating. In the interaction part, the coupling constants $U_{ijkl}$ are composed of the scattering lengths $a_S$ for even total spin $S$ of the scattering particles,

$$U_{ijkl} = \sum_{S=0}^{2F-1} g_S \sum_{M=-S}^{S} \langle ik | SM \rangle \langle SM | jl \rangle \text{ with } g_S = \frac{4\pi\hbar^2 a_S}{m}.$$ We describe the time evolution of the system by first introducing the single-particle density matrix $\rho_{ij}(\boldsymbol{x},\boldsymbol{y},t) = \text{Tr}\left(\hat{\rho}(t)\hat{\psi}_i^\dagger(\boldsymbol{x})\hat{\psi}_j(\boldsymbol{y})\right)$, where $\hat{\rho}(t)$ denotes the statistical many-body density operator whose time evolution is given by $i\hbar\dfrac{d}{dt}\hat{\rho}(t) = \left[\hat{H},\hat{\rho}(t)\right]$. Therefore, the single-particle operator evolves in time according to

$$i\hbar\frac{d}{dt}\rho_{ij}(\boldsymbol{x},\boldsymbol{y}) = \left\langle \left[\hat{\psi}_i^\dagger(\boldsymbol{x})\hat{\psi}_j(\boldsymbol{y}),\hat{H}\right]\right\rangle.$$

When evaluating the right-hand-side of this expression, one encounters quartic expectation values in the interaction part, to which we apply a Hartree-Fock approximation and decompose them as

$$\langle \hat{\psi}_i^\dagger \hat{\psi}_j^\dagger \hat{\psi}_k \hat{\psi}_l \rangle \approx \langle \hat{\psi}_i^\dagger \hat{\psi}_l \rangle \langle \hat{\psi}_j^\dagger \hat{\psi}_k \rangle - \langle \hat{\psi}_i^\dagger \hat{\psi}_k \rangle \langle \hat{\psi}_j^\dagger \hat{\psi}_l \rangle,$$

such that we obtain a closed, but nonlinear equation for $\rho_{mn}(\boldsymbol{x},\boldsymbol{y})$ only, without retaining few-particle density matrices. For a semiclassical treatment, we introduce the Wigner-transform

$$w_{ij}(\boldsymbol{x},\boldsymbol{p}) = \int \frac{d\boldsymbol{y}}{(2\pi\hbar)^3} e^{i\boldsymbol{p}\cdot\boldsymbol{y}/\hbar} \rho_{ij}(\boldsymbol{x}+\boldsymbol{y}/2,\boldsymbol{x}-\boldsymbol{y}/2)$$

of the density matrix, obtaining its phase-space representation, the Wigner function. It is a 10x10 matrix in spin space and a function of position and momentum. A Wigner transform of the entire equation of motion for $\hat{\rho}$ results in an equation of the following form:

$$\partial_t w_{ij}(\boldsymbol{x},\boldsymbol{p}) = \partial_0 w_{ij}(\boldsymbol{x},\boldsymbol{p}) + \frac{q}{i\hbar}(i^2 - j^2) w_{ij}(\boldsymbol{x},\boldsymbol{p})$$

$$+ \sum_{\alpha=0}^{\infty} \frac{1}{\alpha!}\left[\frac{i\hbar}{2}\nabla_y \cdot \nabla_p\right]^\alpha \sum_k \left\{V_{ki}(\boldsymbol{y})w_{jk}(\boldsymbol{x},\boldsymbol{p}) - (-1)^\alpha V_{jk}(\boldsymbol{y})w_{ki}(\boldsymbol{x},\boldsymbol{p})\right\}\bigg|_{\boldsymbol{y}=\boldsymbol{x}}$$

where we have introduced the notation $\partial_0 = -\dfrac{\boldsymbol{p}\cdot\nabla_x}{m} + m\sum_i \omega_i^2 x_i \dfrac{d}{dp_i}$ for the free particle motion in the trap. We also have defined the Hartree-Fock mean-field potential

$$V_{ij}(\boldsymbol{x}) = \sum_{kl} \int d\boldsymbol{p}(U_{jikl} - U_{jlki}) w_{kl}(\boldsymbol{x},\boldsymbol{p}).$$

In the next step, we apply a semiclassical approximation to the orbital degrees of freedom, retaining only the leading term in the gradient expansion above, truncating it at $\alpha = 1$. This is justified as long as the mean-field potential varies slowly compared to the single-particle wavelengths involved. In the general case, one has to include the external fields in this expansion as well, but for a harmonic potential as well as for the quadratic Zeeman effect induced by a homogeneous external magnetic field the truncation is exact. We obtain a semiclassical equation of motion taking the form of a collisionless Boltzmann equation

$$\partial_t w(\boldsymbol{x},\boldsymbol{p}) = \partial_0 w(\boldsymbol{x},\boldsymbol{p}) + \frac{1}{i\hbar}\left[V(\boldsymbol{x}) + qS_z^2, w(\boldsymbol{x},\boldsymbol{p})\right] + \frac{1}{2}\left\{\nabla_x V(\boldsymbol{x}),\cdot\nabla_p w(\boldsymbol{x},\boldsymbol{p})\right\}, \tag{S3}$$

consisting of the kinetic term, a commutator and an anticommutator. The leading term is the commutator of the Wigner function with the mean-field and the Zeeman energy (*31*). This term drives collective and coherent spinor dynamics of the whole Fermi sea. One can easily check that for an initial state without coherences (i.e. without off-diagonal matrix elements in spin space with respect to the quantization axis defined by the external magnetic field) the commutator vanishes and collective spin-changing dynamics is absent. Hence, non-zero off-diagonal elements in the Wigner function are necessary to observe collective spin oscillations. The anticommutator term provides mean-field corrections to the trapping potential. It describes deformations of the total density distribution due to interactions and is much weaker than the commutator in our experiment as reflected in the obtained results. Neglecting this term, we arrive at equation (1) in the main text, apart from the collisional integral $I_{ij}(\boldsymbol{x},\boldsymbol{p})$ worked out in detail in section (S4).

## S3 Single-mode approximation

In this section, we derive a single-mode approximation for a fermionic many-body system, which is widely employed in the main text and describes the observed collective spin-changing behavior very well.

### S3.1 Transformation to the rotating frame

We use a dimensionless description of the many-body system in units of the trapping potential, which is obtained by measuring lengths and momenta in direction $i$ in units of $\sqrt{\hbar/m\omega_i}$ and $\sqrt{m\hbar\omega_i}$, respectively, and by using $\hbar\omega$, $\omega$ and $\omega^{-1}$ as the units for energy, angular velocity, and time. Here $\omega$ can be defined as the geometric mean of the trap frequencies. In these units, equation (S3) reads

$$\left[\partial_t + \sum_{i=1}^{d} \omega_i \left(p_i \partial_{x_i} - x_i \partial_{p_i}\right)\right] w(\boldsymbol{x}, \boldsymbol{p}, t) = \left[V(\boldsymbol{x}, t) + q S_z^2, w(\boldsymbol{x}, \boldsymbol{p}, t)\right]. \tag{S4}$$

While the left-hand side of this equation describes the spin-independent classical motion in the harmonic trap, the right-hand side induces coherent spin dynamics. The latter happens on a much longer timescale than the former. Neglecting the right-hand side for short times, the spin configuration, which initially, at $t=0$, is described by the matrix $w$ at a given point $(\boldsymbol{x}_0, \boldsymbol{p}_0)$ in phase space, is transported along the classical trajectory given by $x_i(t) = \rho_i \cos(\omega_i t - \varphi_i)$ and $p_i(t) = -\rho_i \sin(\omega_i t - \varphi_i)$ with $\rho_i = \left(x_{0i}^2 + p_{0i}^2\right)^{1/2}$ and $\varphi_i = \arctan(p_{0i} / x_{0i})$, i.e. $w(\boldsymbol{x}(t), \boldsymbol{p}(t), t) \approx w(\boldsymbol{x}_0, \boldsymbol{p}_0, 0)$. This motivates a transformation to a reference frame in phase space that is co-moving with the classical dynamics, namely $(\boldsymbol{x}, \boldsymbol{p}) \to (\boldsymbol{x}', \boldsymbol{p}')$ and $g(\boldsymbol{x}, \boldsymbol{p}, t) \to g'(\boldsymbol{x}', \boldsymbol{p}', t) = g(\boldsymbol{x}(\boldsymbol{x}', \boldsymbol{p}', t), \boldsymbol{p}(\boldsymbol{x}', \boldsymbol{p}', t), t)$ for a function $g$ acting in phase space. Here $x_i' = c_i(t) x_i - s_i(t) p_i$ and $p_i' = s_i(t) x_i + c_i(t) p_i$ or $x_i = c_i(t) x_i' + s_i(t) p_i'$ and $p_i = -s_i(t) x_i' + c_i(t) p_i'$ with the short hand notation $c_i(t) = \cos(\omega_i t)$ and $s_i(t) = \sin(\omega_i t)$.

In the new frame, the dynamics is described by

$$\partial_t w'(\boldsymbol{x}', \boldsymbol{p}', t) = \left[V'(\boldsymbol{x}', \boldsymbol{p}', t) + q S_z^2, w(\boldsymbol{x}', \boldsymbol{p}', t)\right].$$

Now the mean-field potential also depends on the new momentum variables and is given by an integral along $d$ rotating axes

$$V'(\boldsymbol{x}', \boldsymbol{p}', t) = U^A \int d^d \boldsymbol{a}\, w'(\boldsymbol{x}' + \boldsymbol{u}(\boldsymbol{a}, t), \boldsymbol{p}' + \boldsymbol{v}(\boldsymbol{a}, t), t) \ .$$

Here $u_i(a_i, t) = -s_i(t) a_i$ and $v_i(a_i, t) = c_i(t) a_i$ and where $U^A w' = \sum_{kl} U^A_{jikl} w'_{kl}$ defines a tensor product with $U^A_{ijkl} = U_{ijkl} - U_{ilkj}$.

### S3.2 Dynamically-induced effective long-range interactions

If the dynamics generated by the commutator is slow compared to the dynamics in the trap, we can approximate the mean-field potential $V'$ by averaging over the rotational dynamics along the different trap directions, as it is captured by $\boldsymbol{u}$ and $\boldsymbol{v}$. For this purpose, we artificially induce a second time $\tau$ describing the dynamics in the trap, while $t$ captures the dynamics induced by the commutator. This allows us to average the mean-field potential with respect to the former. In the following, we distinguish two cases.

### S3.2.1 Commensurate trap frequencies

If the trap frequencies are commensurate, we can average over a time span given by the least common devisor $T_{\text{lcd}}$ of the periods $T_i = 2\pi/\omega_i$ provided that $T_{\text{lcd}}$ is much smaller than the timescale $T_s$ associated with the spin-dynamics induced by the commutator, $T_{\text{lcd}} \ll T_s$. This gives

$$V'_{\text{eff}}(x', p', t) = g \frac{1}{T_{\text{lcd}}} \int_0^{T_{\text{lcd}}} d\tau \int d^d a \, w'(x' + u(a, \tau), p' + v(a, \tau), t).$$

Thus each phase space point $(x', p')$ interacts via the mean-field potential with the $d+1$-dimensional subspace of the 2d-dimensional phase space covered by $(x' + u(a, \tau), p' + v(a, \tau))$ during integration. The most elementary examples are given by an isotropic trap with $\omega_i = 1$ and $T_{\text{lcd}} = 2\pi$ or by the one-dimensional system. For the latter, we can write

$$\frac{1}{2\pi} \int_0^{2\pi} d\tau \int_{-\infty}^{\infty} da = \frac{1}{\pi} \int du \int dv \left(u^2 + v^2\right)^{-1/2}$$ yielding the effective long-range interaction potential in phase space $U^A \pi^{-1} \left( \left(x^{(1)} - x^{(2)}\right)^2 + \left(p^{(1)} - p^{(2)}\right)^2 \right)^{-1/2}$ between two particles 1 and 2, introduced in (*28*).

### S3.2.2 Incommensurate trap frequencies

If the trap frequencies are pairwise incommensurate such that $T_{\text{lcd}} \to \infty$ or simply if $T_{\text{lcd}}$ becomes large compared to the individual periods $T_i$, we can still approximate the mean-field potential by an effective time-averaged one. However, in this case the double separation of timescales $\max(T_i) \ll T_{\text{av}} \ll T_s$ is required, with $T_{\text{av}}$ denoting the time-interval over which the average shall be taken. This interval has to be large compared to the individual periods $T_i$, such that $T_{\text{av}} \approx n_i T_i$ with integers $n_i \gg 1$. Excluding resonance effects between the dynamics along the different trap directions one can, in a second step, assume independent averages along the individual trap directions

$$V'_{\text{eff}}(x', p', t) = U^A \frac{1}{T_{\text{av}}} \int_0^{T_{\text{av}}} d\tau \int da \, w'(x' + u(a,\tau), p' + v(a,\tau), t)$$

$$\approx U^A \frac{1}{2\pi} \int_0^{2\pi} d\varphi_1 \int_{-\infty}^{\infty} da_1 \cdots \frac{1}{2\pi} \int_0^{2\pi} d\varphi_d \int_{-\infty}^{\infty} da_d \, w'(x' + u(a,\tau), p' + v(a,\tau), t)$$

$$\approx U^A \int du \int dv \left( \prod_{i=1}^{d} \frac{\pi^{-1}}{\sqrt{u_i^2 + v_i^2}} \right) w'(x' + u, p' + v, t).$$

Thus, we arrive at a mean-field potential corresponding to an anisotropic long-range interaction

$$U^A \sum_{i=1}^{d} \pi^{-1} \left( \left( x_i^{(1)} - x_i^{(2)} \right)^2 + \left( p_i^{(1)} - p_i^{(2)} \right)^2 \right)^{-1/2}$$ between two particles 1 and 2 in phase space.

### S3.2.3 Description in single-mode approximation

One effect of the dynamically induced effective long-range interaction is that it smoothens the mean-field interaction potential with respect to the density distribution $n(x) = \int dp \, w(\mathbf{x}, \mathbf{p})$ in the trap. For an initial state $w_{ij}(x, p, t) = \rho_{ij}(t) f(x, p)$, characterized by a spin configuration described by $\rho_{kl}$, which is homogeneous in phase space, and by a phase-space distribution $f(x, p) = f(\{x_i^2 + p_i^2\})$, which is invariant under the dynamics in the trap. This has the consequence that a dephasing of the spin dynamics in phase space resulting from mean-field interactions is suppressed by the dynamics in the trap. Hence we might approximate the state of the system by

$$w_{kl}(x, p, t) \approx \rho_{kl}(t) f(x, p)$$

also for later times.

### S3.3 Comparison of the single-mode approximation with experimental data and full 1d theory

We have thoroughly compared the experimentally obtained data with numerical results obtained from a single-mode approximation and full numerical calculations with momentum and spatial degrees of freedom in one dimension.

For the single-mode approximation, we assume a fixed spatial and momentum distribution justifying a separation ansatz of the form $w_{ij}(x, p, t) = \rho_{ij}(t) \cdot f(x, p)$, where we derive $f(x, p)$ from the fermionic statistics. Plugging this into equation (S4) leads to the following equation accounting for the full spin-9/2 system:

$$\partial_t \rho_{ij}(t) = \frac{\int d^3x \, n^2(x)}{i\hbar \cdot N} \left[ \sum_{kl} (U_{klij} - U_{kjil}) \cdot \rho_{kl}(t), \rho(t) \right]_{ij} + \left[ qS_z^2, \rho(t) \right]_{ij} \tag{S5}$$

with $n(x) = \int dp \, f(x, p)$. We additionally include a damping term of the form $-\Gamma \cdot \rho_{ij}(t)$ for $i \neq j$ with the damping rate $\Gamma$ accounting for the experimentally observed damping. To compare the single-mode approximation with the experiments, we insert a damping rate of $\Gamma = 2\,\text{Hz}$ for the simulations in Fig. 2 and of $\Gamma = 1\,\text{Hz}$ for the simulations in Fig. 3. Solving equation (S5) numerically, we extract frequencies with a Fourier analysis and amplitudes by taking the difference between maximum and minimum of the first oscillation. The simulations are performed without free parameters and the results are plotted in Figs. 2A, 2B, and 3B. We checked, that frequency and amplitude do not depend on the phenomenologically implemented damping rates. Since in our experiments the spin oscillations mainly occur between the spin configuration $\pm 1/2$ and $\pm 3/2$, we average the results of the Fourier analysis between these components and always depict the normalized frequency spectrum for each magnetic field, each density and each temperature, respectively. The frequencies are calculated for the mean experimental parameters and reveal no significant deviations including experimental uncertainties. The depicted amplitudes are also averaged between the participating spin configurations $\pm 1/2$ and $\pm 3/2$. For the magnetic field dependence depicted in Fig. 2A, the influence of the experimental uncertainties on the numerical results for the amplitude is also small and hence not shown. For the simulation of the density and temperature dependence, we take into account experimental uncertainties, namely temperature fluctuations in Fig. 2B (ranging from $0.1\,T_f$ to $0.3\,T_f$) and density fluctuations in Fig. 3B (ranging from $4.9 \times 10^{12}\,\text{cm}^{-3}$ to $5.6 \times 10^{12}\,\text{cm}^{-3}$), reflected in the corresponding shaded areas.

Since the time-of-flight measurements mix momentum and spatial components in our experiments, these data lack the full spatial information and hence the experimental evidence of suppression of spatial deformations, in particular if these are small. To check the validity of the single-mode approximation in more detail, we employed an in-situ detection protocol, where we use microwave pulses at $1.3\,\text{GHz}$ to transfer all except one spin component to the $F = 7/2$ manifold, which is off-resonant to the detection light. After the time evolution, we switch off all potentials for $1\,\text{ms}$ to detect the spatial structure of the atomic sample. We do not observe any formation of spatial structures.

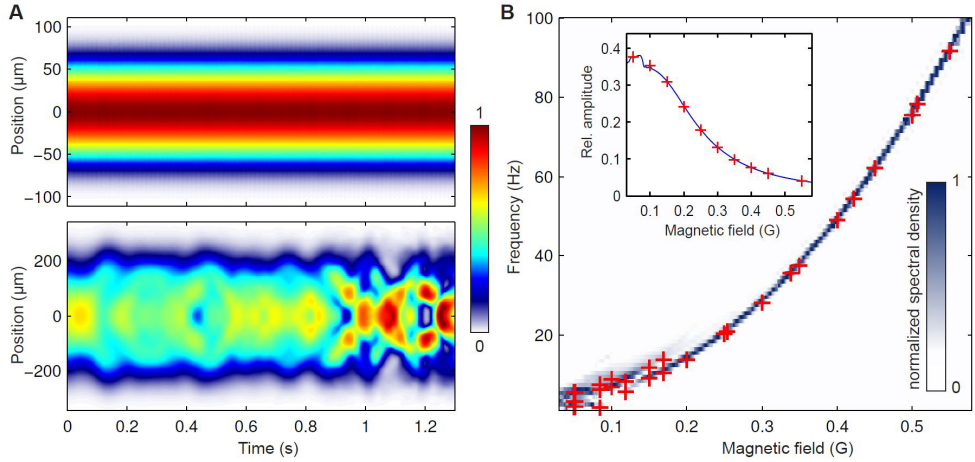

**Fig. S2. Validity of the single-mode approximation.**
**A** Numerical simulation of the time evolution of the normalized spatial distribution of the spin component $m=1/2$ in two different regimes: The trap-dominated regime (upper graph), where $\omega_{\text{trap}} = 2\pi \times (32,43,137)\,\text{Hz}$ is larger than the interaction-driven spin-changing oscillations and the single-mode approximation is valid, and the interaction-dominated regime (lower graph), where $\omega_{\text{trap}} = 2\pi \times (5,200,200)\,\text{Hz}$ is comparable to the interaction-driven spin-changing oscillations and the single-mode approximation fails. In both cases, the used parameters are $B=0.12\,\text{G}$, $N=1.4\times 10^5$ and $T=0.13\,T_{\text{f}}$. **B** Simulation of the frequency (main graph) and amplitude (inset) as a function of the applied magnetic field. The experimental parameters from the measurements in Fig. 2A are used. Solid lines in insets correspond to the amplitude and the shading in main graph to a Fourier spectrum, both deduced from calculations within a single-mode approximation. Crosses are the extracted amplitudes and frequencies from calculations with full momentum and spatial resolution in one dimension. Note that at lower magnetic fields, several frequencies significantly contribute to the dynamics.

For a better understanding of the validity of the single-mode approximation, we compared the single-mode calculations with simulations including full momentum and spatial resolution in one dimension (while treating the other two dimensions in single-mode approximation) for a wide range of different parameters. In the regime where our experiments were performed, we find an excellent agreement of the single-mode approximation with the full calculations. The upper graph of Fig. S2A exemplarily shows the time evolution of the normalized spatial distribution of the spin component $m=1/2$, which exhibits large-amplitude spin oscillations normalized in this graph, but clearly the emergence of any spatial deformations is suppressed. In this simulation, parameters have been used where the kinetic motion induced by the trap is faster than the interaction-driven spin-changing dynamics, as it is the case in our experiments.

Numerically, we can also access a regime, where the kinetic motion induced by the trap and the interaction-driven spin-changing dynamics occur on comparable timescales. The result of the corresponding normalized time evolution in this regime is exemplarily plotted in the lower graph in Fig. S2A. We find the intriguing formation of pronounced spatial structures, clearly revealing the breakdown of the single-mode approximation. Note that this regime is experimentally not accessible in our system. To ensure the validity of the single-mode approximations even further for the parameters, at which the experiments have been performed, we also compare the single-mode results with the full calculations for different magnetic fields where we choose the same parameters. Fig. S2B shows the extracted frequencies and amplitudes as a function of the magnetic field for both calculations revealing perfect agreement. We conclude that the single-mode approximation is valid within our parameter regime as already shown in the experiments in the main text.

## S4 Incoherent collisions in the relaxation approximation

In this section, we deduce the collisional integral $I_{ij}(\boldsymbol{x},\boldsymbol{p})$ and the temperature-dependent damping rate within the relaxation approximation.

To estimate the damping rate we derive an explicit expression for the collision term. We follow the approach of Ref. (*31*), where a Boltzmann equation is derived from the T-matrix describing a single scattering event of two particles. This approach can be generalized to arbitrary spin by introducing a T-matrix for each channel defined by the total spin $S$ of both particles:

$$\langle i,k|\hat{T}_q|j,l\rangle = \sum_{SM}\langle ik|SM\rangle\langle SM|jl\rangle T_S(q) \text{ with } T_S(q) = \frac{4\pi\hbar^2}{m}a_S(1-iqa_S)+O(a_S^3)$$

with Clebsch-Gordan coefficients $\langle SM|jl\rangle$ and the wave number $q$ characterizing the energy of the relative motion of both particles. The starting point is an expansion in powers of the scattering lengths $a_S$. The lowest order, linear in $a_S$, gives rise to the mean-field interaction terms which have been derived in a different way already above. The mean-field interaction corresponds to the forward scattering fraction of the T-matrix: $\langle i,k|\hat{T}_q|j,l\rangle = U_{ijkl}+O(a_S^2)$.

The terms quadratic in $a_S$, which describe lateral collisions where particles can exchange momentum, contribute to the collision term. Such collisions appear as decoherence in a description formulated in terms of the single-particle density matrix and are not captured by a mean-field potential. They describe relaxation processes or particle-hole excitations in the Fermi sea. We expect these collisions to be the most important mechanism for the damping of coherent spin oscillations visible in the experimental results.

A way to obtain a damping rate is to use a linearized version of the Boltzmann equation and the collision integral. This means to decompose the Wigner function like $w_{ij}(\boldsymbol{x},\boldsymbol{p},t) = w_{ij}^0(\boldsymbol{x},\boldsymbol{p}) + \delta w_{ij}(\boldsymbol{x},\boldsymbol{p},t)$ into a stationary part and a small time-dependent deviation. We assume the stationary part to be of the form $w_{ij}^0(\boldsymbol{x},\boldsymbol{p}) = f_0(\boldsymbol{x},\boldsymbol{p}) M_{ij}$, a product of the equilibrium phase space distribution for a two-component Fermi gas

$$f_0(\boldsymbol{x},\boldsymbol{p}) = \frac{1}{(2\pi\hbar)^3} \left( \exp\frac{1}{k_B T_0}\left[\frac{p^2}{2m} + V^{\text{trap}}(\boldsymbol{x}) - \mu_0 \right] + 1 \right)^{-1}$$

with a matrix $M_{ij}$ created by the preparatory radio-frequency pulse. Note that for the initial two-component system with a total number of $N$ atoms in the trap, we have $\int d\boldsymbol{x} \int d\boldsymbol{p} f_0(\boldsymbol{x},\boldsymbol{p}) = \frac{N}{2}$. This is a good approximation for the experimentally prepared initial state, thus the theory linearized about this state should be valid for short times. The experimental results (Figs. 2c and 3a in the main text) show that after some time the system reaches a pre-equilibrium state, where coherent spin oscillations involving the $m = \pm 1/2, \pm 3/2$ components have decayed, whereas a significant population of the other spin states, eventually found in equilibrium, has not yet been established. We assume that the pre-equilibrium state can be approximated by an equilibrium phase-space distribution

$$f_1(\boldsymbol{x},\boldsymbol{p}) = \frac{1}{(2\pi\hbar)^3} \left( \exp\frac{1}{k_B T_1}\left[\frac{p^2}{2m} + V^{\text{trap}}(\boldsymbol{x}) - \mu_1 \right] + 1 \right)^{-1},$$

where we choose $T_1$ and $\mu_1$ such that the total energy and particle number is conserved when four components are occupied, e.g. $\int d\boldsymbol{x} \int d\boldsymbol{p} f_1(\boldsymbol{x},\boldsymbol{p}) = \frac{N}{4}$, assuming approximately equal population of the components $m = (-3/2, -1/2, 1/2, 3/2)$ while neglecting the population in all other spin states. To account for the decay from the initial state $w_{ij}^0(\boldsymbol{x},\boldsymbol{p})$ towards a state described by $f_1$, we make the simple ansatz $\delta w_{ij}(\mathbf{x},\mathbf{p},t) = \left(f_0(\mathbf{x},\mathbf{p}) - f_1(\mathbf{x},\mathbf{p})\right) A_{ij}(t)$, capturing the impact of two-particle scattering in a minimal fashion. With $A_{ij}(t) = \frac{4}{N}\int d\mathbf{x}\int d\mathbf{p}\, \delta w_{ij}(\mathbf{x},\mathbf{p},t)$ the time evolution in tensor notation is given by

$$\frac{d}{dt}A - \frac{1}{i\hbar}\Omega_{\text{MF}}(A) = -\Gamma A + I[w_0],$$

where $\Omega_{\text{MF}}$ represents the coherent oscillatory mean-field contribution while the terms on the right arise from the linearized collision integral. $I[w_0]$ contains the time-independent part of the collision term and is quadratic in $w_0$ and the tensor $\Gamma$ contains the damping rates for each matrix element of the Wigner function. For the thermal relaxation rate we are interested in the temperature dependence of $\Gamma$, which is found to be spin-independent:

$$\Gamma_{ijkl} = \gamma_{ijkl} I(k_B T) \tag{S6}$$

with $I(k_B T) = \int d\mathbf{x}\int d\mathbf{p}\int d\mathbf{q}\, |\mathbf{p}+\mathbf{q}|\, f_0(\mathbf{x},\mathbf{p})\left(f_0(\mathbf{x},\mathbf{q}) - f_1(\mathbf{x},\mathbf{q})\right)$.

In Fig. 3C in the main text, we have plotted the experimentally observed value of $\Gamma$ in comparison to the calculated damping of the Wigner function component $w_{3/2,3/2}$. The calculations agree well at low temperatures but its increase for high temperatures remains below the experimentally measured values.

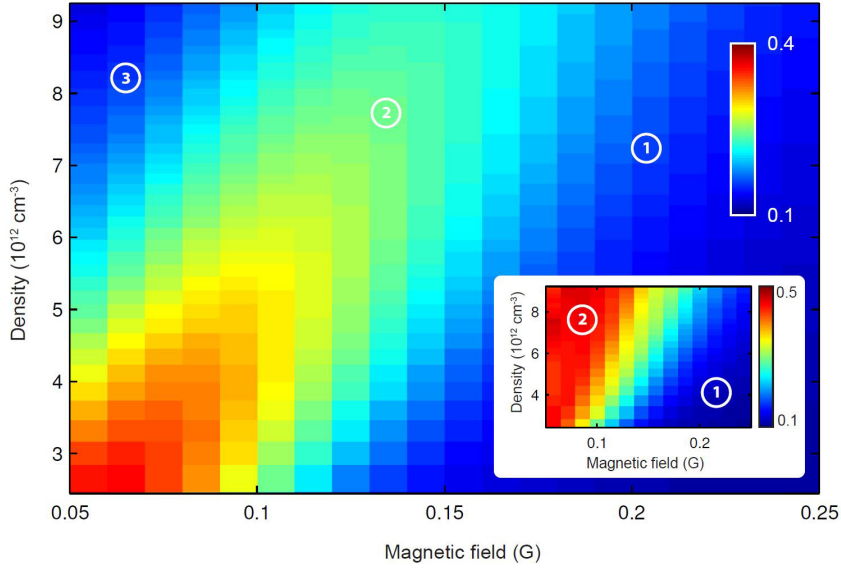

**Fig. S3. Numerical simulation of the stability diagram.**

Shown is the calculated spin occupation $n(\pm 1/2)$ after a time evolution of $3.5\,\mathrm{s}$, depending on magnetic field and density. We take the parameters from the experiments in Fig. 4 and use as the initial state spin occupations $n(\pm 3/2)$ prepared with a spin rotation angle of $\vartheta = 0.1$ to account for the small seed. The simulations are performed within the single-mode approximation, whereas the damping rate is calculated within the relaxation approximation, scaled with a global factor of $2.25$. We can well reproduce the different regions of the stability diagram: a Zeeman-protected regime ①, a spin-oscillation regime ② and the collisionally-stabilized regime ③. In the inset, a stability diagram with a weak constant damping rate of $2\,\mathrm{Hz}$ is plotted. Only the Zeeman-protected regime ① and the spin-oscillation regime ② are found, whereas the collisionally-stabilized regime is absent.

Combining the single-mode approximation and the relaxation approximation, which describe the coherent mean-field-driven spin dynamics and its damping, respectively, we simulate the stability diagram shown in Fig. 4 in the main text. For these calculations, we use the experimental parameters as in Fig. 4 and approximate the initial seed with a small spin rotation angle of $\vartheta = 0.1$, applied to the initial spin mixture $m_1, m_2 = \pm 3/2$. We include in the single-mode calculations ad-hoc a linear damping term characterized by a damping rate $\Gamma$, which is taken from the relaxation approximation using equation (S6). We include in the single-mode calculations a damping term with the damping rate $\Gamma$, which is calculated within the relaxation approximation using equation (S6). This rate increases with density, which is crucial for the collisionally-stabilized regime. We find reasonable agreement with our experimental data using the calculated damping rate scaled with a global factor of $2.25$ as depicted in Fig. S3. The simulation reproduces the smooth transition from the Zeeman-protected regime ① to the spin-oscillation regime ②. In addition, it also reproduces the collisionally-stabilized regime ③ at high densities and low magnetic fields. To illustrate the interplay between the coherent dynamics and the damping rate in more detail, a calculation of the stability diagram for a rather weak density-independent damping rate of $2\,\text{Hz}$ is depicted in the inset of Fig. S3. This simulation also reproduces the transition from the Zeeman-protected regime ① to the spin-oscillation regime ②. However, the stabilized regime at high densities is absent, which underlines that the stabilization mechanism is a "spin-Zeno" effect: At larger densities, increased incoherent collisions "project" the spins back onto the initial state and thus suppress the coherent evolution, stabilizing the system in its magnetically-excited spin configuration.